\documentclass[aps,prb,twocolumn,superscriptaddress,floatfix]{revtex4}
\usepackage{color}
\usepackage{graphicx}
\usepackage{epsfig}
\usepackage{textcmds}
\bibstyle{apsrev.bib}
\usepackage{amsmath,amssymb}
\usepackage{graphicx}

\begin{document}
\input epsf.sty

\title{Nonequilibrium dynamics of fully frustrated Ising models at $T=0$}

\author{M. Karsai}
\affiliation{
Institute of Theoretical Physics,
Szeged University, H-6720 Szeged, Hungary}
\affiliation{Institut N\'eel-MCBT
 CNRS 
\thanks{U.P.R. 5001 du CNRS, Laboratoire conventionn\'e
avec l'Universit\'e Joseph Fourier}, B. P. 166, F-38042 Grenoble,
France}
\author{J-Ch. Angl\`es d'Auriac}
\affiliation{Institut N\'eel-MCBT
 CNRS 
\thanks{U.P.R. 5001 du CNRS, Laboratoire conventionn\'e
avec l'Universit\'e Joseph Fourier}, B. P. 166, F-38042 Grenoble,
France}
\affiliation{Laboratoire de Physique Subatomique et de Cosmologie 53, av. des Martyrs F-30026 Grenoble,France}
\author{F. Igl\'oi}
\affiliation{
Research Institute for Solid State Physics and Optics,
H-1525 Budapest, P.O.Box 49, Hungary}
\affiliation{
Institute of Theoretical Physics,
Szeged University, H-6720 Szeged, Hungary}

\date{\today}

\begin{abstract}
We consider two fully frustrated Ising models: the antiferromagnetic triangular model in a field of strength, $h=H T k_B$, as well as
the Villain model on the square lattice. After a quench from a disordered initial state to $T=0$ we study the nonequilibrium dynamics 
of both models by Monte Carlo simulations. In
a finite system of linear size, $L$, we define and measure sample dependent ``first passage time'', $t_r$,
which is the number of Monte Carlo steps until the energy is relaxed to the ground-state value. The
distribution of $t_r$, in particular its mean value, $\langle t_r(L) \rangle$, is shown to obey the scaling
relation, $\langle t_r(L) \rangle \sim L^2 \ln(L/L_0)$, for both models.  Scaling of the autocorrelation function of the antiferromagnetic triangular model
is shown to involve logarithmic corrections, both at $H=0$ and at the field-induced Kosterlitz-Thouless
transition, however the autocorrelation exponent is found to be $H$ dependent.
\end{abstract}

\maketitle

\newcommand{\bc}{\begin{center}}
\newcommand{\ec}{\end{center}}
\newcommand{\be}{\begin{equation}}
\newcommand{\ee}{\end{equation}}
\newcommand{\beqn}{\begin{eqnarray}}
\newcommand{\eeqn}{\end{eqnarray}}

\section{Introduction}
\label{sec:Intr}
The phenomenon of ageing in physics is first noticed in glassy systems\cite{Bouchaud,Cugliandolo}, when the physical properties
of materials are found to depend on the preparation process. In a microscopic point of view during
relaxation these systems goes over an infinite set of metastable states and thermodynamic equilibrium is
never reached in finite time. Later it was observed that similar phenomenon could take place in
simple homogeneous ferromagnets\cite{Bray}, too, if the the system is quenched from the paramagnetic phase below or at the
critical temperature, $T_c$. After the quench domains of parallel spins are started to grow and the typical
size of a domain of correlated sites, $\xi(t)$, after time $t$ is given by
\be
\xi(t) \sim t^{1/z}
\label{t_z}
\ee
$z$ being the dynamical exponent. In the case of a critical quench generally $z$ corresponds to the
dynamical exponent measured at equilibrium and it generally does not depend on the temperature of the initial state, $T_i$.
This is, however not true for the two-dimensional $XY$ model in which case at the two sides of the Kosterlitz-Thouless temperature, $T_{KT}$,
we have\cite{Bray00}:
\be
\xi(t) \sim
\begin{cases}
t^{1/2}, &\text{if $T_i<T_{KT}$},\\
(t/\ln t)^{1/2}, &\text{if $T_i>T_{KT}$}.
\label{XY}
\end{cases}
\ee
Due to the quench the translational symmetry in
time is broken consequently the autocorrelation function, $G(t,s)$, measured after a waiting time $s<t$ is
non-stationary. In the limit $t,s \gg 1$ it is expected to behave as\cite{Janssen89,Godreche02,Calabrese05}:
\be
G(t,s) \simeq s^{-a} \tilde{G}\left(\frac{t}{s}\right)\;,
\label{auto}
\ee
whereas for the $XY$ model the scaling function depends on the combination\cite{Bray00}, $y=t\ln s/s \ln t$:
\be
G(t,s) \simeq s^{-a} \tilde{G}\left(\frac{t \ln s}{s \ln t}\right)\;.
\label{autoXY}
\ee
In both cases the exponent, $a$, is related to the decay exponent of critical spatial correlations:
\be
a=\frac{d-2+\eta}{z}\;,
\label{a}
\ee
and the scaling function, $\tilde{G}(y)$, for large arguments behaves as\cite{huse89}:
\be
\tilde{G}(y) \sim y^{-\lambda/z}\;,
\label{lambda}
\ee
where the autocorrelation exponent, $\lambda$, is a new parameter which can not be expressed in terms of other (equilibrium)
exponents.

There is a class of systems in which due to competing interactions or frustration the critical point is shifted to $T_c=0$,
thus there is no ordered phase. The prototype of such systems are the antiferromagnetic triangular Ising model\cite{Wannier,Houtappel} (ATIM) and
the Villain model\cite{Villain}, which is the fully frustrated Ising model (FFIM) on the square lattice. Both models have a finite
ground state entropy per site and the usual ensemble average for $T>0$ is replaced by averaging over the different degenerate
ground state configurations. It is known exactly for several decades that the ATIM and the FFIM are in the same universality
class\cite{Stephenson,Forgacs,Peschel,Zittartz} as far as the decay of the spin-spin correlation function is considered. 
In the presence of a weak magnetic field of strength, $h=H T k_B$, the ATIM is shown to stay critical\cite{Blote91} for
$H \le H_c$. At the end of the critical line at $H=H_c$ there is a Kosterlitz-Thouless transition\cite{Blote93}, which is analogous to that
observed in the $XY$ model.

As far as the nonequilibrium critical properties of the ATIM and the FFIM after a quench to $T_c=0$ are concerned they are also expected to be in
the same universality class, but about the value of the exponents, such as $z$ and $\lambda$, there are conflicting
results in the literature, which all refer to $H=0$. In Ref.\cite{Moore99} the authors suggest that in the ATIM there are topological defects
which are attracted or repelled by entropy-driven Coulomb forces, which are analogous to vortices and anti-vortices in the two-dimensional $XY$
model for $T>T_{KT}$. Therefore nonequilibrium relaxation in
the two models are expected to be governed by the same mechanism and nonequilibrium scaling is given in the same form, see the second equation of (\ref{XY}). 
Indeed numerical calculation of the density of topological defects are in favour of these considerations.
On the other hand, in Ref.\cite{Kim} the numerical data are found in favour of a distinct dynamical exponent, $z \simeq 2.33$, which is
explained in terms of interaction between defects and loose spins. In this study the autocorrelation function is found
to involve the exponent, $\lambda/z \simeq 0.86$. More recently the nonequilibrium dynamics of the FFIM after a quench to $T_c=0$
is studied numerically\cite{Walter} and the obtained data are interpreted in favour of the same type of scaling as in the $XY$ model for $T>T_{KT}$, i.e. with $z=2$
supplemented by logarithmic corrections.

We note on studies of related models which can be obtained through mappings to two-dimensional covering problems\cite{Blote82,Blote84}, such as by rhombi or dominoes or
by dimers in the dual lattices. As described in section \ref{sec:Model} these are also equivalent to solid-on-solid (SOS) models. Recently
it has been shown\cite{blunt} that the p-terphenyl-3,5,3',5'-tetracarboxylic acid adsorbed on graphite self-assembles into a two-dimensional rhombus random tiling.
Its relaxation dynamics and relation with the similar process in structural glasses is studied in Ref\cite{garrahan}. The relaxation
dynamics of the compressible ATIM, in which also elastic strain fields are taken into account is studied in Ref.\cite{chakraborty}, whereas the ATIM in a
staggered magnetic field is considered in Ref\cite{dhar}.

In this paper we study the nonequilibrium critical dynamics of the ATIM and the FFIM by Monte Carlo (MC) simulations and try to solve the
existing controversy about their scaling properties. Our approach differs from the previous ones in two aspects. First, in a
finite system of linear size, $L$, we define and measure a quantity, the sample (or initial condition) dependent first passage time, $t_r$.
Its distribution and in particular its mean value, $\langle t_r(L) \rangle$, is used to obtain a direct relation between the
characteristic time and the length in the system and through Eq.(\ref{t_z}) we obtain a direct estimate for the dynamical exponent.
We note that in previous numerical studies the calculated critical quantities are characterised by combination of different critical
exponents, see e.g. the autocorrelation function in Eq.(\ref{auto}), and no direct estimate for $z$ has been made. The second feature
of our study is that for the ATIM we consider not only the zero field model, but the Kosterlitz-Thouless transition at $H=H_c$ as well. We
compare the numerical results in the two cases in order to decide if the two transitions are in the same dynamical universality class.

The structure of the paper is the following. In Sec.\ref{sec:Model} the models are described. In Sec.\ref{sec:Relax} we present
results about the sample dependent first passage times both for the ATIM (at $H=0$ and at $H=H_c$) and for the FFIM. The autocorrelation
function of the ATIM is studied in Sec.\ref{sec:Auto}. We close our paper by a discussion.

\section{Models}
\label{sec:Model}
We consider two fully frustrated Ising models. The first model is the antiferromagnetic Ising model on the triangular lattice in the presence of an external field, $h$,
which is in the order of the thermal fluctuations, $\sim k_B T$. The Hamiltonian is defined in terms of $\sigma_i= \pm 1$ as
\be
\frac{{\cal H}}{k_B T}=-K \sum_{\langle i,j \rangle} \sigma_i \sigma_j - H \sum_i \sigma_i\;,
\label{Hamilton}
\ee
where $i$ and $j$ are sites of a triangular lattice and the first sum runs over nearest neighbours. We consider the limit
$T \to 0$, so that $K \to -\infty$ but at the same time we have $H=O(1)$.

The model is exactly solved at zero field, when the entropy per site is\cite{Wannier,Houtappel}:
\be
{\cal S}=\frac{2}{\pi} \int_0^{\pi/3} \ln(2 \cos \omega) {\rm d} \omega \simeq 0.32306\;,
\label{Entropyformula}
\ee
which is calculated for periodic boundary conditions. We note that the actual form of the boundary
conditions is crucial for the ground state entropy. For example considering a hexagonal
part of the lattice with open boundary conditions the entropy can 
be any value in the range $\left[0,\frac{3}{2}\ln 3 - 2\ln 2\right]$ depending on the ratio
of the lengths of the sample\cite{Elser}. On the other hand for a part with parallelogram shape
and open boundary conditions there are only two ground states and consequently the entropy is zero\cite{Millane}.
The spin-spin correlation function at $H=0$ is calculated by Stephenson\cite{Stephenson} which shows algebraic decay: $r^{-1/2}$, thus $\eta=1/2$. 

Further interesting results are obtained through different mappings, which are valid also for non-equal couplings as well as in the presence of
external fields. The model is equivalent to a covering problem with rhombi in the triangular lattice or with dimer covering in the dual hexagonal
lattice\cite{Blote82,Blote84}. Assigning appropriate height variables to the rhombi one obtains a solid-on-solid (SOS) model which has a roughening transition at
$H=H_c$ of the Kosterlitz-Thouless type and the rough phase is present for small fields. The SOS model is then transformed by a standard although non-exact mapping
to a Gaussian model\cite{Blote91,Blote93} from which the conformal anomaly is obtained as $c=1$ and different
spin- and vertex-exponents are calculated at $H=0$. For non-zero field the model is studied numerically by transfer-matrix techniques\cite{Blote93}.
The system is shown to be critical for $0 \le H \le H_c$ and has long range order otherwise.
The location of the Kosterlitz-Thouless transition is found
at $H_c \simeq 0.266$. Some exponents are found to be constant along the critical line, whereas others are $H$-dependent but they satisfy
the weak-universality hypothesis, i.e. the ratio of specific exponents is constant. At the Kosterlitz-Thouless transition point the spin-spin
correlation function is conjectured to decay with an exponent: $\eta_c=4/9$.

The second model is the fully frustrated Ising model on the square lattice defined by the Hamiltonian:
\be
\frac{{\cal H}}{k_B T}=-\sum_{\langle i,j \rangle} K_{i,j}\sigma_i \sigma_j.
\label{Hamilton1}
\ee
In the even vertical lines the nearest neighbour couplings are antiferromagnetic, $K_{i,j}=-K$, otherwise they are ferromagnetic, $K_{i,j}=K$.
Consequently at each elementary square the product of the couplings have a negative sign, and the external field is $H=0$. This model can
be mapped to a domino covering problem in the square lattice\cite{Villain} or equivalently to a dimer covering problem in the dual lattice\cite{fisher,kasteleyn}. From this
mapping one can obtain the ground state entropy per site\cite{fisher,Villain,Andre}:
\be
{\cal S}=\frac{G}{\pi}=0.2916
\ee
where $G$ is the Catalan's constant. 
The spin-spin correlation function is calculated either through a mapping to the eight-vertex model\cite{Forgacs} or by transfer matrix
techniques\cite{Peschel,Zittartz}. The decay exponent is found to be  $\eta=1/2$, which is the same as for the ATIM  at $H=0$. In a similar
way as the ATIM, the FFIM is mapped to a SOS model on the square lattice\cite{henley}. Analysing the relaxation process of this model in equilibrium
a lower bound for the relaxation time is obtained: $\tau(L) \ge O(L^2)$, thus $z \ge 2$. This relation in the continuum approximation turns
to an equality.

\section{Relaxation time in finite systems}
\label{sec:Relax}

We study the models by MC simulations. They are prepared in a random initial state (i.e. with $T=\infty$) and then quenched to $T_c=0$ and let
to evolve according to the Glauber dynamics. A randomly chosen spin at site $i$ is flipped with a probability: $p(i)=e^{-\Delta E_i/k_B T}/
(e^{\Delta E_i/k_B T}+e^{-\Delta E_i/k_B T})$, where the reduced energy-difference is: $-\Delta E_i/k_B T=\sigma_i(K\sum_j' \sigma_j+H)$
and the sum in $j$ runs over nearest neighbours of $i$. In the limit we consider ($K \to -\infty$) a flip is always accepted if the
interaction energy is lowered ($1^{st}$ type flip) and accepted with a probability: $p_i=e^{-H \sigma_i}/(e^{H \sigma_i}+e^{-H \sigma_i})$, for loose spins,
i.e. for which $\sum_j' \sigma_j=0$ ($2^{nd}$ type flip). 

During the relaxation process in an infinite system the interaction energy is
monotonously decreasing due to $1^{st}$ type flips. In a finite sample, however, at time $t_r$ the interaction energy reaches its
minimum, corresponding to a ground-state configuration of the $H=0$ problem. For later steps, $t>t_r$, only $2^{nd}$ type flips are
possible and the system evolves in the ground-state sector, i.e. the dynamics is in equilibrium.
Consequently for a given initial state (sample) $t_r$ is a characteristic time
that we call a first passage time, and we assume that it scales as a usual relaxation time. Here
we use periodic boundary conditions. For open boundaries possibly one should modify the definition of $t_r$, c.f. measuring
the time until the energy is lowered by a given fraction, but the same qualitative picture is expected as for periodic lattices.

In the following of this section we study the distribution and properties
of $t_r$ for different finite sizes, $L$. Note that the Monte-Carlo algorithm is not ergodic
and does not sample properly the ground state properties \cite{Kim}, it is merely a single spin flip dynamic. 
We expect that the results we present below will remain unchanged if we use another dynamic, as soon as
the number of spins which can be flip at a time is kept fixed when $L$ varies.

In the actual calculation, as already mentioned in section \ref{sec:Model}, we consider finite systems with $L^2$ sites, with periodic boundary conditions
and with $L$, which is a
multiple of 3. We have checked that the asymptotic dependence of $t_r$ with $L$ is the same if $L=3l+1$
or $L=3l+2$, but with an increased prefactor and therefore the computation time is also larger. 
The ground states are usually classified using the notion of strings (see Ref. \cite{dhar}), but 
the dynamic always drives the system into the most probable string sector.

\begin{figure}[h]
\centerline{\epsfxsize=3.25in\ \epsfbox{
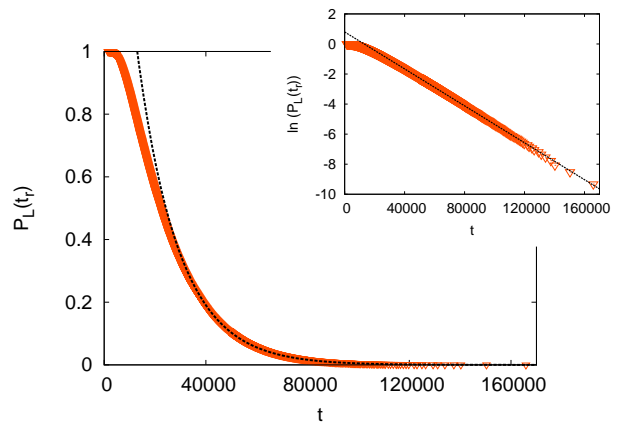}
}
\caption{(Color online) Accumulated distribution of the sample dependent first passage time, $t_r$, for the ATIM 
at $H=0$ with linear size, $L=192$.
For large arguments the tail of the distribution is exponential, as shown in the inset.
   }
\label{Abb1}
\end{figure}

We went up to $L=768$ and the number of realizations are given in Table \ref{table:1} for the different models.
As an example the accumulated distribution of $t_r$, $P_L(t_r)$, for $L=192$
is shown in Fig.\ref{Abb1} for the ATIM at $H=0$.
%%%%%%%%%%%%%%%%%%%%%%%%%%%%%%%%%%%%%%%%%%%FIG 1.%%%%%%%%%%%%%%%%%%%%%%%%%%%%%%%%%%%%%%%%%%%%%%%%%%%%%%

%%%%%%%%%%%%%%%%%%%%%%%%%%%%%%%%%%%%%%%%%%%%%%%%%%%%%%%%%%%%%%%%%%%%%%%%%%%%%%%%%%%%%%%%%%%%%%%%%%%%%%%

\begin{figure}[h]
\centerline{\epsfxsize=3.25in\ \epsfbox{
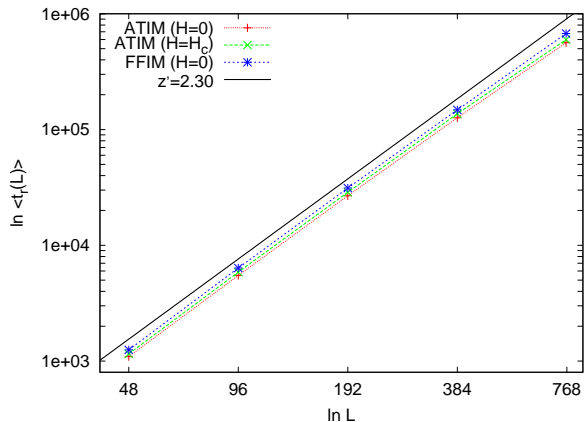}
}
\caption{(Color online) Average first passage time, $\langle t_r(L) \rangle$, as a function of $L$ in a log-log plot for the three frustrated models studied in the paper. The straight full line has a slope $z'=2.30$ obtained by simple linear
fit. Note that for large $L$ the local slope between the points is smaller than $z'$, see also Table \ref{table:1}.
}
\label{Abb2}
\end{figure}

The tail of $P_L(t_r)$ for large argument is exponential, $P_L(t_r) \sim \exp(-t_r/\tau(L))$, as shown in the inset of Fig.\ref{Abb1}
in a lin-log plot. For a given size, $L$, we have calculated the mean value, $\langle t_r(L) \rangle$, which is presented in Table \ref{table:1}. We have checked, that the
ratio $\langle t_r(L) \rangle/\tau(L)$ is approximately independent of $L$, thus there is just one time-scale in the problem.
To see the size-dependence of the time-scale we have plotted in Fig.\ref{Abb2}  $\langle t_r(L) \rangle$  vs. $L$ in a double-logarithmic scale for the three different problems
we studied.
%%%%%%%%%%% FIG 2  %%%%%%%%%%%%%%%%%%%%%%%%%%%%%%%
%
%
%%%%%%%%%%%%%%%%%%%%%%%%%%%%%%%%%%%%%%%%%%%%%%

In all cases the points are approximately
on parallel straight lines, which would correspond to the relation in Eq.(\ref{t_z}) by substituting $\xi$ by $L$. Having a simple linear fit through the points
we obtain $z' \approx 2.30$, which is in agreement with the estimate of $z=2.33$ found in Ref.\cite{Kim} for the ATIM with $H=0$. However, having a closer look at the
points in Fig.\ref{Abb2} on can see that the local slopes of the curve is monotonously decreasing with increasing $L$.
We have calculated effective exponents, $z_{eff}(L)$, from two-point fit comparing results with $L$ and $L'$:
\be
z_{eff}(L)=\frac{\ln\langle t_r(L) \rangle-\ln\langle t_r(L') \rangle}{\ln(L)-\ln(L')}\;.
\label{zeff}
\ee
which are presented in Table \ref{table:1}, too.

For a given size, $L$, within the error of the calculation the $z_{eff}(L)$-s are found to be the same for the three problems and they
are decreasing below $z'=2.30$.  We have extrapolated them by assuming
a logarithmic correction in the form of $1/\ln(L/L_0)$, where $L_0$ is an $O(1)$ constant.

%%%%%%%%%%%%%%%%%%%%%%%%%%%%%%%%%%%%%%%%%%%FIG 3.%%%%%%%%%%%%%%%%%%%%%%%%%%%%%%%%%%%%%%%%%%%%%%%%%%%%%%

\begin{figure}[h]
\centerline{\epsfxsize=3.25in\ \epsfbox{
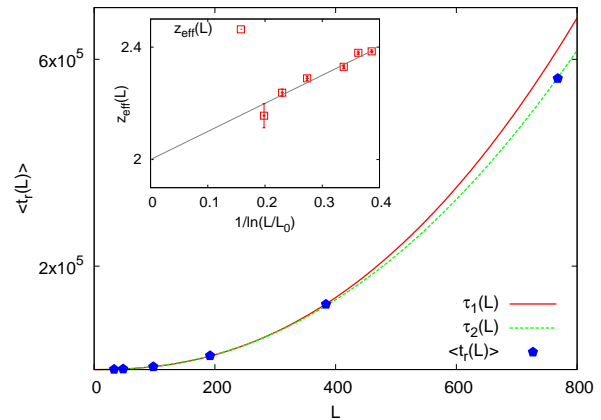}
}
\caption{(Color online) Fitting the measured average first passage times, $\langle t_r(L) \rangle$, of the ATIM at $H=0$ by the two possible functional
forms in Eqs.(\ref{fit1}) and (\ref{fit2}). In the inset extrapolation of the effective dynamical exponents in Table \ref{table:1}
are given as a function of $1/\ln(L/L_0)$ with $L_0=2.5$.
   }
\label{Abb3}
\end{figure}

%%%%%%%%%%%%%%%%%%%%%%%%%%%%%%%%%%%%%%%%%%%%%%%%%%%%%%%%%%%%%%%%%%%%%%%%%%%%%%%%%%%%%%%%%%%%%%%%%%%%%%%

This extrapolation is shown in the inset of Fig.\ref{Abb3} for
the ATIM at $H=0$. The effective exponents for different $L$ lay on a straight line of slope $1$ and the extrapolated value is close to $2$, which is in agreement with the
scenario in Ref.\cite{Moore99}, i.e. logarithmic corrections to the $z=2$ leading behaviour, as given for the $XY$-model in Eq.(\ref{XY}). Indeed inverting the second equation
of (\ref{XY}) we have (if $\xi$ is replaced by $L$) $t_r \sim L^2 \ln(L)$, from which the effective exponent is given in leading order as: $z_{eff}(L) \simeq 2 +1/\ln(L/L_0)$.

We have also explicitly confronted the two possible scenarios:
\be
\tau_1(L)=A L^z
\label{fit1}
\ee
and
\be
\tau_2(L)=A L^z \ln(L/L_0)
\label{fit2}
\ee
with the data $\langle t_r(L) \rangle$ in Table \ref{table:1}. We have minimised the quantity
\be
\sum_L \left|\frac{\tau_k(L)-\langle t_r(L) \rangle}{\tau_k(L)+\langle t_r(L) \rangle} \right|
\label{mini}
\ee
for $k=1$ and $2$ and in the sum $L$ runs over the considered sizes. For the ATIM with $H=0$ the best fit in the two cases are obtained with:
$\tau_1(L)=0.18 L^{2.26}$, and $\tau_2(L)=0.183 L^2 \ln(L/3.6)$, which are shown in the main part of Fig.\ref{Abb3}. Indeed, the second scenario
yields a much better fit and note that in this case the value of the dynamical exponent, $z=2$, is obtained from the optimisation process, its
value was not fixed in advance. The value of the quantity in Eq.(\ref{mini}) is slowly varying if $z$ is in the range $\left[1.96,2.04\right]$ with a minimum
located precisely at $z=2$, outside this range, we observe a rapid increase. Repeating this analysis for the two other frustrated models 
we have obtained the same conclusion. For the ATIM at $H=H_c$
the best fits are $\tau_1(L)=0.18 L^{2.28}$, and $\tau_2(L)=0.20 L^2 \ln(L/3.9)$, whereas for the FFIM $\tau_1(L)=0.20 L^{2.27}$, and $\tau_2(L)=0.22 L^2 \ln(L/4.1)$.
In both cases the relative error of the fits were 5 to 10-times smaller for the second scenario.
\begin{widetext}
\begin{center}
\begin{table}[h]
\caption{Average of the sample dependent first passage time $\langle t_r \rangle$ for different finite systems of the
ATIM at zero field and at the field induced Kosterlitz-Thouless transition point, $H=H_c$, as well as for
the FFIM at zero field. The effective dynamical exponent calculated from two consecutive sizes, see Eq.(\ref{zeff}) is
also indicated together with the number of samples used in the simulation. \label{table:1}}
\begin{tabular}{|c|c|c|r|c|c|r|c|c|r|}  \hline
&\multicolumn{3}{|c|}{ATIM($H=0$)}&\multicolumn{3}{|c|}{ATIM($H=H_c$)}& \multicolumn{3}{|c|}{FFIM}\\ \hline
L&$log_2\langle t_r \rangle$ & $z_{eff}$ & samples & $log_2\langle t_r \rangle$ & $z_{eff}$ &samples& $log_2\langle t_r \rangle$ & $z_{eff}$ &samples\\ \hline
33  & 8.80  & 2.38 & 200000 & & & & & &  \\
48  & 10.09 & 2.33 & 200000 & 10.17 & 2.35 & 200000 & 10.29 & 2.35 & 192000 \\
96  & 12.42 & 2.29 & 200000 & 12.52 & 2.29 & 200000 & 12.64 & 2.29 &  96000\\
192 & 14.71 & 2.24 & 89848  & 14.81 & 2.25 & 103799 & 14.93 & 2.24 &  57000\\
384 & 16.95 & 2.15 & 35271  & 17.06 & 2.13 & 38414  & 17.17 & 2.19 &  31612\\
768 & 19.10 &      & 1667   & 19.19 &      & 1701   & 19.37 &      &  3157\\ \hline
\end{tabular}
\end{table}
\end{center}
\end{widetext}
\section{Autocorrelation function of the ATIM}
\label{sec:Auto}
For the FFIM the scaling behaviour of the autocorrelation function in the ageing regime has been thoroughly studied in Ref.\cite{Walter}. In that work the numerical
data for $G(t,s)$ has been analysed by assuming two possible scaling forms, as given in Eqs.(\ref{auto}) and (\ref{autoXY}) and a clear evidence
in favour of the second scenario is found. To summarise for the FFIM the scaling function is suggested to contain logarithmic corrections and the autocorrelation
exponent is measured $\lambda=d$. Here we repeat this type of analysis for the AFIM, both at $H=0$ and at $H=H_c$.

\subsection{ATIM at $H=0$}

%%%%%%%%%%%%%%%%%%%%%%%%%%%%%%%%%%%%%%%%%%%FIG 4.%%%%%%%%%%%%%%%%%%%%%%%%%%%%%%%%%%%%%%%%%%%%%%%%%%%%%%

\begin{figure}[h]
\centerline{\epsfxsize=3.25in\ \epsfbox{
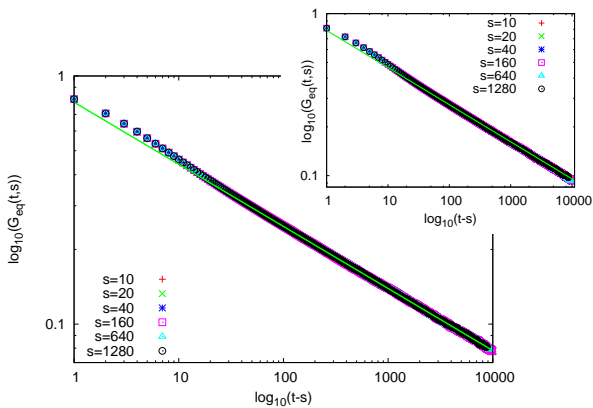}
}
\caption{(Color online) The equilibrium autocorrelation function, $G_{eq}(t,s)$ as a function of $t-s$ for the ATIM at $H=0$ in a log-log plot. The asymptotic slope of
the curve is given by $a(H=0)=0.2501(10)$. Inset: the same analysis at $H=H_c$. The asymptotic slope of
the curve is given by $a(H=H_c)=0.229(5)$.}
\label{Abb4a}
\end{figure}

%%%%%%%%%%%%%%%%%%%%%%%%%%%%%%%%%%%%%%%%%%%%%%%%%%%%%%%%%%%%%%%%%%%%%%%%%%%%%%%%%%%%%%%%%%%%%%%%%%%%%%%

%%%%%%%%%%%%%%%%%%%%%%%%%%%%%%%%%%%%%%%%%%%FIG 5.%%%%%%%%%%%%%%%%%%%%%%%%%%%%%%%%%%%%%%%%%%%%%%%%%%%%%%

\begin{figure}[h]
\centerline{\epsfxsize=3.25in\ \epsfbox{
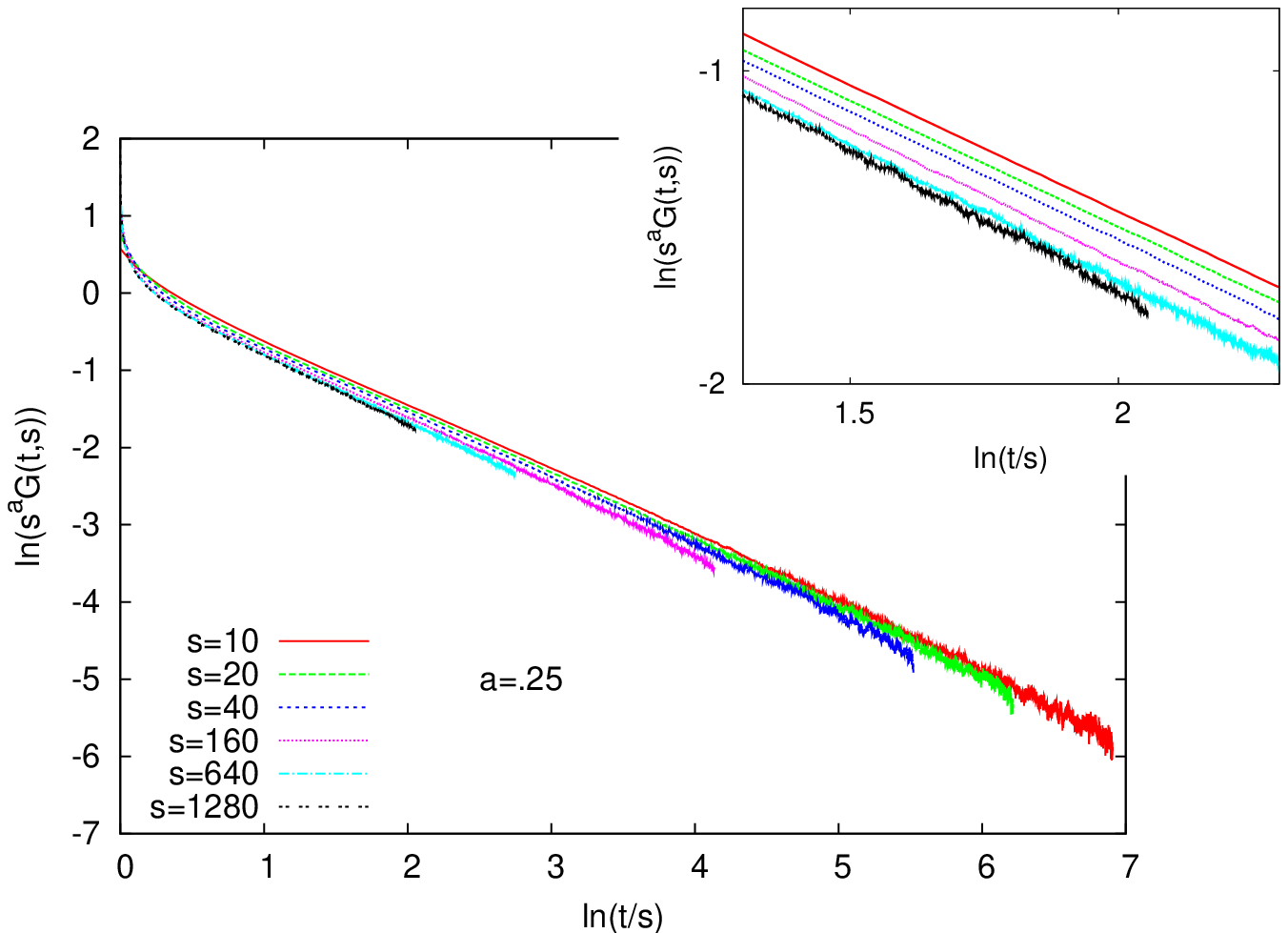}
}
\caption{(Color online) Scaling of the reduced autocorrelation function, $\tilde{G}(t,s)=s^a G(t,s)$ of the ATIM at $H=0$ with $a=0.25$ assuming conventional scaling, see
Eq.(\ref{auto}).}
\label{Abb4}
\end{figure}

%%%%%%%%%%%%%%%%%%%%%%%%%%%%%%%%%%%%%%%%%%%%%%%%%%%%%%%%%%%%%%%%%%%%%%%%%%%%%%%%%%%%%%%%%%%%%%%%%%%%%%%

%%%%%%%%%%%%%%%%%%%%%%%%%%%%%%%%%%%%%%%%%%%FIG 6.%%%%%%%%%%%%%%%%%%%%%%%%%%%%%%%%%%%%%%%%%%%%%%%%%%%%%%

\begin{figure}[h]
\centerline{\epsfxsize=3.25in\ \epsfbox{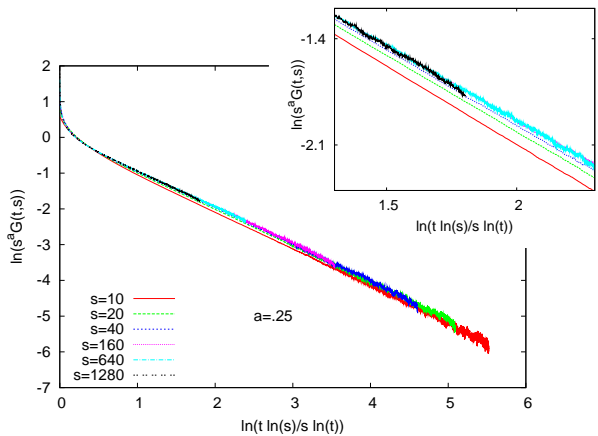
}
}
\caption{(Color online) The same as in Fig. \ref{Abb4} with $XY$-type scaling, see Eq.(\ref{autoXY}).
   }
\label{Abb5}
\end{figure}

%%%%%%%%%%%%%%%%%%%%%%%%%%%%%%%%%%%%%%%%%%%%%%%%%%%%%%%%%%%%%%%%%%%%%%%%%%%%%%%%%%%%%%%%%%%%%%%%%%%%%%%

First we consider the zero field model and analyse the autocorrelation function at equilibrium, i.e. formally the measurement is performed for $t,s > t_r$.
In the actual calculation we have started from a ground state and after thermalization we have performed the measurement.
In that case due to time translational invariance $G_{eq}(t,s)$ is stationary and we have asymptotically (i.e. for $t-s \gg 1$):
\be
G_{eq}(t,s) \approx (t-s)^{-a}\;.
\ee
This behavior is illustrated in Fig. \ref{Abb4a}  in which the decay exponent is measured as $a(H=0)=0.2501(10)$, in agreement with the scaling relation in Eq.(\ref{a}).

Next we consider the autocorrelation function in the ageing regime, i.e. for $t,s < t_r$ and analyse the
scaling function, $\tilde{G}(t,s)=s^a G(t,s)$, assuming the two scenarios in Eqs.(\ref{auto}) and (\ref{autoXY}).
For the decay exponent we used the presumably exact value, $a=0.25$.
For conventional scaling, i.e. with $y=t/s$, as given in Eq.(\ref{auto}) the results are shown in Fig.\ref{Abb4} , whereas for $XY$-type scaling., i.e. with $y=t \ln s/s \ln t$, as given in Eq.(\ref{autoXY}) the results are in Fig.\ref{Abb5}. As seen in Fig.\ref{Abb4} for conventional scaling the collapse of the curves for different values of the waiting times, $s$, is
certainly not perfect. In the enlarged part of the figure presented in the inset is clearly seen a shift between the curves and this shift can be approximately
characterised by $\ln s$. On the other hand in Fig.\ref{Abb5} for $XY$-type scaling the collapse of the curves is much better: for the three largest waiting times: $s=160, 640$
and $1280$ the curves practically fall into each other. Consequently our data are in favour of logarithmic corrections in the scaling function of the autocorrelation
function of the ATIM at $H=0$. We have then estimated the autocorrelation exponent, $\lambda/z$, from the slope of the linear part of Fig.\ref{Abb5} for the three largest waiting times
and we obtained: $\lambda/z=1.003(10)$. That value corresponds to $\lambda=d$, as obtained for the FFIM\cite{Walter}.

%%%%%%%%%%%%%%%%%%%%%%%%%%%%%%%%%%%%%%%%%%%FIG 7.%%%%%%%%%%%%%%%%%%%%%%%%%%%%%%%%%%%%%%%%%%%%%%%%%%%%%%

\begin{figure}[h]
\centerline{\epsfxsize=3.25in\ \epsfbox{
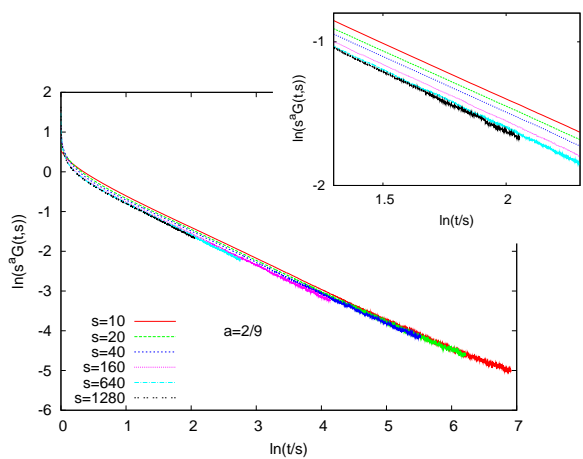}
}
\caption{(Color online) Scaling of the reduced autocorrelation function, $\tilde{G}(t,s)=s^a G(t,s)$ of the ATIM at $H=H_c$ with $a=2/9$ assuming conventional scaling, see
Eq.(\ref{auto}).}
\label{Abb6}
\end{figure}

%%%%%%%%%%%%%%%%%%%%%%%%%%%%%%%%%%%%%%%%%%%%%%%%%%%%%%%%%%%%%%%%%%%%%%%%%%%%%%%%%%%%%%%%%%%%%%%%%%%%%%%

%%%%%%%%%%%%%%%%%%%%%%%%%%%%%%%%%%%%%%%%%%%FIG 8.%%%%%%%%%%%%%%%%%%%%%%%%%%%%%%%%%%%%%%%%%%%%%%%%%%%%%%

\begin{figure}[h]
\centerline{\epsfxsize=3.25in\ \epsfbox{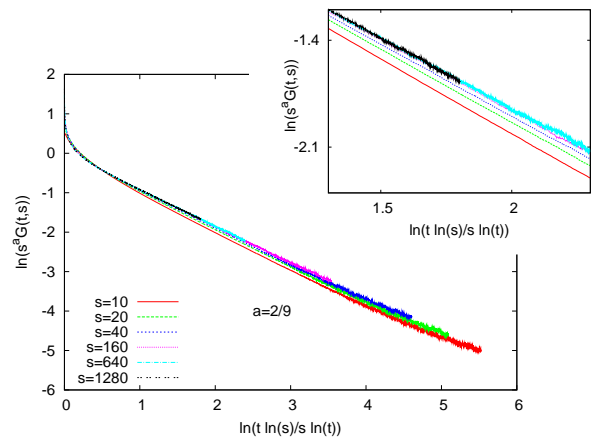
}
}
\caption{(Color online) The same as in Fig. \ref{Abb6} with $XY$-type scaling, see Eq.(\ref{autoXY}).
   }
\label{Abb7}
\end{figure}

%%%%%%%%%%%%%%%%%%%%%%%%%%%%%%%%%%%%%%%%%%%%%%%%%%%%%%%%%%%%%%%%%%%%%%%%%%%%%%%%%%%%%%%%%%%%%%%%%%%%%%%

\subsection{ATIM at $H=H_c$}

We have repeated the analysis of the autocorrelation function at the field-induced Kosterlitz-Thouless transition point, at $H=H_c$. In that case the equilibrium
autocorrelation function is shown in the inset of Fig.\ref{Abb4a}, in which case we have measured the decay exponent as $a(H=H_c)=0.229(5)$, close to
the conjecturedly exact value: $\eta_{c}/z=2/9$. Results about the autocorrelation scaling function are shown in Fig.\ref{Abb6} for conventional scaling and in 
Fig.\ref{Abb7} for $XY$-type scaling. As for
the zero field case the scaling collapse of the curves for different values of the waiting time are much better for $XY$-type scaling. We have also measured
the autocorrelation exponent with the result: $\lambda/z=0.93(2)$. This is somewhat smaller than those found in the other two models.

\section{Discussion}
\label{sec:Disc}
We have studied the nonequilibrium dynamics of different fully frustrated Ising models in which the critical
point is at $T=0$. Two of these models, the ATIM at zero field and the FFIM are in the same universality class, as far as
equilibrium critical properties are concerned, for example the decay exponent is $\eta=1/2$ for both models. The third model, the ATIM at the critical field, $H=H_c$, 
has a Kosterlitz-Thouless transition and also the correlation exponent is different. We have quenched the models from a disordered initial state to the
critical temperature, i.e. to $T=0$, and studied the nonequilibrium domain growth process as well as the scaling properties of the autocorrelation function.

The domain growth is studied at a fixed size of the system, $L$, by measuring the time, $t_r$, until equilibrium is reached. Although $t_r$ varies
from sample to sample its distribution is characterised by only a single time-scale, which can be taken as the average value , $\langle t_r \rangle$. For all the three models
$\langle t_r \rangle$ is found to have the same asymptotic $L$ dependence, which is in the form known for the $XY$-model in Eq.(\ref{XY}). Since both the
$XY$-model and the ATIM at $H=H_c$ have the same Kosterlitz-Thouless transition this equivalence found in the non-equilibrium critical dynamics is not surprising.

Using this result we can estimate the asymptotic behaviour of the density of defects, $\rho(t)$, in the systems. To be concrete we consider the ATIM at $H=0$
for which a defect is represented by an elementary triangle having all the three spins in the same state. In a finite sample of linear size, $L$, and sample
dependent first passage time, $t_r$, at $t=t_r$ there is one (or $O(1)$) defect, 
consequently the density of defects at $t=\langle t_r(L) \rangle$ is $\rho(t) \sim L^{-2}$.
Using the relation in Eq.(\ref{XY}) and replacing $\xi$ by $L$ we arrive to the result: $\rho(t) \sim \log(t/t_0)/t$, in agreement with the conjecture in
Ref.\cite{Moore99}.

The analogy with the nonequilibrium scaling of the $XY$-model is found to be valid for the autocorrelation function, too. For the ATIM, both at zero field and
at $H=H_c$ optimal scaling collapse of the autocorrelation function is found if the scaling variable is taken in the same form, as in the $XY$-model. For the
ATIM at $H=0$ and in the FFIM the autocorrelation exponent is found to be $\lambda=d$, like in mean-field theory, however with logarithmic corrections.
On the other hand in the ATIM at the field-induced Kosterlitz-Thouless transition point the autocorrelation exponent is found somewhat smaller, than $d$.
Consequently in the ATIM with varying field the dynamical exponent is presumably constant, $z=2$, the scaling functions contain same type of logarithmic
corrections, but the autocorrelation function weakly depends on the value of $H$.

We thank for useful discussions with C. Chatelain and J-C. Walter as well as for useful comments by an
anonymous Referee.
This work has been supported by the German-Hungarian exchange program (DFG-MTA) and by the
Hungarian National Research Fund under grant No OTKA TO48721, K62588, K75324.
M.K. thanks the Minist\`ere Fran\c{c}ais des
Affaires \'Etrang\`eres for a research grant.

\end{document}